\documentclass[12pt]{iopart}

\begin{document}

\title[Unitary decoupling treatment of a quadratic bimodal \textbf{CQED} model]{Unitary decoupling treatment of a quadratic bimodal \textbf{CQED} model}

\author{M. Guccione$^{ 1}$,  M. A. Jivulescu$^{2}$, A. Messina$^1$}

\address{$^1$  Dipartimento di Fisica, Universit\`{a} di Palermo, via Archirafi
36, 90123 Palermo, Italy\\
$^2$ Department of
Mathematics,  University " Politehnica" of Timi\c{s}oara,
P-ta Victoriei Nr. 2,
 300006 Timi\c{s}oara,
Romania}
\ead{marina.guccione@unipa.it, antonino.messina@unipa.it, maria.jivulescu@mat.upt.ro}
\begin{abstract}
We consider a  two-photon quantum model of radiation-matter interaction between a single two-level atom and a degenerate bimodal high-Q cavity field. Within this tripartite system the explicit construction of two collective radiation modes, one of which freely evolving and the other one quadratically coupled to the matter subsystem, is reported. Meaning and advantages of such a decoupling treatment are carefully discussed.
\end{abstract}

\maketitle

\section{Introduction}

Investigating on the physical properties of confined quantum matter-radiation systems has been, is and shall be always on fashion. This topic indeed meets the increasing demand of new more and more miniaturized devices for industrial applications which, in turn, spurs research activities on fundamental issues of quantum mechanics. Understanding the nano-world physical behaviour speeds up the development of current technologies and provides radically new ideas for applicative advances.
Many theoretically envisaged and experimentally realized physical situations wherein a matter-radiation system lives in a micro or nano confinement have originated successful new branches of research as for example Cavity Quantum Electrodynamics (CQED) \cite{II4:15}, \cite{II4:16}, \cite{II4:17}, \cite{I4:6}, \cite{I4:7}, \cite{I4:8}, \cite{I4:9}, \cite{I4:10}, Circuit Quantum Electrodynamics(Circuit QED) \cite{II5:9}, \cite{II4:2}, \cite{II4:3}, \cite{II4:4} Trapped Atoms \cite{II5:7}, \cite{II5:8}, \cite{III4} and Quantum Dots \cite{II5:9}, \cite{II5:10}.
Of course what plays the role of \textquotedblleft matter\textquotedblright and \textquotedblleft radiation \textquotedblright depends indeed on the specific scenario.
In the simplest case the \textquotedblleft matter\textquotedblright subsystem is generally represented by few-level atoms which might be either true flying or trapped atoms or artificial ones like two-level quantum dots or charge or flux qubit. The radiation subsystem instead is generally represented by one or two quantized bosonic modes which belong  either to an high-Q electromagnetic cavity in a typical CQED situation \cite{II4:15}, to on chip resonators acting as quantized cavities in the circuit QED realm \cite{II4}, \cite{II4:28}, \cite{III2} or to a photonic nanostructure embedding a single quantum dot \cite{II5:2}.
A common aspect characterizing these exemplary four different scenarios is the achievement of effective interactions on demand resulting from the possibility of an high experimental control of relevant physical parameters. This fact paves the way for theoretical investigations and experimental realizations of appropriately tailored new regimes of matter-radiation interaction.
On the basis of the previous considerations it is not surprising that the dynamical evolution of a confined system is often based on appropriate generalizations of the seminal Jaynes-Cummings Model(JCM) \cite{III3:19}. In such extended models the bosonic variables couple in an effective way, sometimes nonlinearly, to fermionic variables describing the internal degrees of freedom of the matter subsystem \cite{II8}, \cite{164}, \cite{173}.
In this paper we are interested in a model describing the two-photon interaction between an effective two-level atom and two quantized modes of a bimodal high-Q degenerate cavity \cite{161} (degenerate means that its two modes involved in the processes possess the same frequency).
Compared with the typical CQED setup including a single mode in a high-Q superconductive cavity, the utilization of bimodal cavities \cite{I4:11} enables experimental investigations on tripartite systems. The aim is to drive and to reveal the appearance of quantum correlations \cite{I4:15}, \cite{I4:12}, \cite{I4:14} which might be useful also for implementing quantum information protocols \cite{I4:12}, \cite{I4:13}. This explains the growing theoretical and experimental interest toward such systems \cite{III2}, \cite{I4:12}, \cite{178}, \cite{I8}, \cite{I10}, \cite{II3}, \cite{175}, \cite{I4}, \cite{15}, \cite{III1}. It is of relevance that quite recently a maximally entangled state of two modes of a high-Q cavity has been experimentally demonstrated  \cite{I4:11}, \cite{I8}.
 The scope of this paper is to show that the Hamiltonian model of bimodal degenerate cavity resonantly coupled by two-photon processes to a single two-level atom is unitarily equivalent to a generalized exactly solvable quadratic JCM and a further collective 
quantized free evolving mode.
Besides being interesting in its own, such a transformation puts at our disposal a simple conceptual tool to improve the physical interpretation of the well-known rich dynamical behavior exhibited by this tripartite CQED system \cite{159}. 
 It has been demonstrated that the time evolution is indeed dominated by a peculiar sensitivity to the parity of the initial population assigned to the Fock state of one of the two modes, the other one and the atom starting from their respective ground states \cite{161}. Such a behaviour is a genuine quantum phenomenon which has been called parity effect \cite{159}, \cite{169}, \cite{168}, \cite{172}.
In the next sections we report our original unitary decoupling treatment leading to a transformed Hamiltonian wherein the two-level atom is effectively (quadratically) coupled to only one out of the two collective cavity modes.
The analysis of dynamical properties of the original physical model associable to the existence of such a decoupling will be reported elsewhere.

\section{The Hamiltonian model}
The one-photon interaction between a two-level atom and two degenerate radiation modes of a lossless high-Q cavity is described by the following hamiltonian \cite{175}
\begin{eqnarray}\label{ham1}H=&\hbar\omega_0 S_z+\hbar\omega\sum\limits_{\mu=1}^2 \alpha_{\mu}^{+}\alpha_{\mu}+\sum\limits_{\mu=1}^2(\hbar g_{\mu}\alpha_{\mu}^{+}S_{-}+\hbar g_{\mu}^{*}\alpha_{\mu}S_{+})
\end{eqnarray}
The energy separation of the two atomic levels $|+\rangle$ and $|-\rangle$ is $\hbar\omega_{0}$ and the pseudospin operators $S_{z}$ and $S_{\pm}$, describing the interval degrees of freedom, are such that
\begin{equation}\label{2}S_{z}|\pm\rangle=\pm\frac{1}{2}|\pm\rangle, S_{\pm}|\pm\rangle=\frac{1}{2}|\mp\rangle
\end{equation}
The $\mu$-th radiation mode is described by Bose operators $\alpha_{\mu}$ and $\alpha_{\mu}^{+}$. No atom-cavity resonance condition is assumed. It is well known that the dynamical behaviour of this extended JCM exhibits a rich variety of attractive quantum effects stemming from the assumption of cavity degeneracy. In particular when at $t=0$ one of the two modes is coherent and the other one and the atom are in their respective ground states, the field dynamics is indeed dominated by oscillations with exchange of a large number of photons between the two modes, displaying collapses and revivals in the field populations. The physical origin of this quantum interference effect may be traced back to the circumstance that in the cavity-atom system two collective modes emerge. What is peculiar is that one of them is coupled to two-level atom in accordance with a JCM characterized by an effective coupling constant, whereas the other one evolves freely. Recently much attention has been devoted to quadratic atom-high-Q cavity interaction models since it has been demonstrated that they predict remarkable quantum features in the dynamics of transparent observables easily experimentally accessible \cite{161}, \cite{168}.
In the trapped ion context \cite{post1}, \cite{post2}, \cite{post3}, \cite{post4}, the same effect is at the origin of a robust parity dependent entanglement between the vibrational and the electronic subsystem \cite{III4:33}, \cite{III4:34}, \cite{III4:35}, \cite{III4:36}. In this paper we focus our attention on the hamiltonian model first introduced in ref\cite{161}:
\begin{eqnarray}\label{ham}H_{eff}=&\hbar\omega_0 S_z+\hbar\omega\sum\limits_{\mu=1}^2 \alpha_{\mu}^{+}\alpha_{\mu}+
sS_z\sum\limits_{\mu=1}^2 \alpha_{\mu}^{+}\alpha_{\mu}+[(r_1\alpha_1\alpha_2^{+}+r_2\alpha_1\alpha_2^{+}S_z)+h.c.]
\nonumber\\
&+[(\sum\limits_{\mu=1}^2\lambda_{\mu} \alpha_{\mu}^2)+g\alpha_1\alpha_2)S_{+}+h.c.]
\end{eqnarray}
describing, in the dipole approximation, the physical behaviour of a single effective two-level atom coupled by two-photon processes to a bimodal, degenerate, high-Q cavity field. The two modes share the same angular frequency $\omega$ and differ either by the polarization vector or by the direction of propagation. This hamiltonian is an effective model reflecting all the energy exchange channels among the three constitutive subsystems under the two-photon resonance condition $\omega_{0}=2\omega$. Thus $g$ is the complex (in general) coupling constant relative to atomic transitions accompanied by a simultaneous gain or loss of one photon in both modes. 
The terms controlled by the real parameters $r_1$ and $r_2$ are Rayleigh terms while the ones involving $s$ (real) are called Stark terms.
 Finally the complex, in general, parameter $\lambda_{\mu}$ with $\mu=1,2$ is associated to two-photon exchange processes between the atom and the $\mu-th$ cavity mode. The time evolution of the tripartite system represented by the nonlinear hamiltonian $H_{eff}$ displays the previously quoted parity effect. This phenomenon consists in an intermode net transfer of photons which turns out to be sensitive to the parity of the Fock population $n(0)$ initially injected in one of the modes. More in detail, after a rapid reduction to 50\% of $n(0)$ and a relatively longer time interval characterized by a negligible intermode exchange of photons, a sudden inversion of the dynamical role of the two modes occurs if $n(0)$ is even. On the contrary, if $n(0)$ is odd, the initially empty mode gives back its adsorbed 50\% of the initial population to the other mode. The physical origin of this phenomenon is traceable back to quantum interference manifestations. As for the case of the linear hamiltonian model given by eq. (\ref{ham1}) where the unitary reduction leads to a deep understanding of the origin of the quantum features dominating the dynamics of the atom-mode-mode system, we wonder whether an analogous treatment of the quadratic hamiltonian model may contribute to a more profound insight of the parity effect and its peculiar consequences.
  This approach amounts at looking for the existence of a unitary operator accomplishing the canonical transformation of $H_{eff}$ to a new fictious system describable as a quadratic interaction between the two-level atom and a collective mode (generalized quadratic JCM) and a second free evolving collective mode. The scope of this paper is to investigate this question providing a conditioned positive reply.

\section{The canonical transformation of $H_{eff}$}
To this end we introduce the following two-parameter transformation:
 \begin{equation}\label{5}\left\{\begin{array}{rl}
&\widetilde{\alpha}_1=(\cos^2\frac{\theta}{2}+e^{-i\eta}\sin^2\frac{\theta}{2} )\alpha_1+\frac{1}{2}\sin\theta(1-e^{-i\eta})\alpha_2\equiv V^{+}\alpha_1V
\\&\widetilde{\alpha}_2=\frac{1}{2}\sin\theta(e^{-i\eta}-1)\alpha_1-(e^{-i\eta}\cos^2\frac{\theta}{2}+\sin^2\frac{\theta}{2} )\alpha_2\equiv V^{+}\alpha_2V
\\&\widetilde{S}_i=S_i\equiv V^{+}S_iV,\quad i=+,-,z
\end{array}\right. .\end{equation}
where $V$ is the unitary operator accomplishing it, certainly existing since it is straightforward to show that whatever the real parameters $\theta$  and $\eta$ are the given transformation expressed by eq. (\ref{5}) is canonical. 
Even if eqs. (\ref{5}) are enough to transform $H_{eff}$, we wish however to construct $V$
and to this end we consider the following hermitian operator
\begin{equation}\label{6}K=\Omega_1\alpha_1^{+}\alpha_1+\Omega_2\alpha_2^{+}\alpha_2+\lambda(\alpha_1^{+}\alpha_2+\alpha_1\alpha_2^{+})
\end{equation}
where the adimensional parameters $\Omega_1$, $\Omega_2$ and $\lambda$ are real. It is immediate to demonstrate that the unitary operator
\begin{equation}\label{7}S(\theta)=\exp[\theta(\alpha_1^{+}\alpha_2-\alpha_1\alpha_2^{+})]
\end{equation}
transforms $K$ as follows

\begin{equation}\label{8}\widetilde{K}:=S^{+}KS=\widetilde{\Omega}_1\alpha_1^{+}\alpha_1+\widetilde{\Omega}_2\alpha_2^{+}\alpha_2+\widetilde{\lambda}(\alpha_1^{+}\alpha_2+\alpha_1\alpha_2^{+})
\end{equation}
where
\begin{equation}\left\{\begin{array}{rl}
&\widetilde{\Omega}_1=\Omega_1\cos^2\theta+\Omega_2\sin^2\theta-\lambda\sin 2\theta
\\& \widetilde{\Omega}_2=\Omega_1\sin^2\theta+\Omega_2\cos^2\theta+\lambda\sin 2\theta
\\ &\widetilde{\lambda}=\frac{1}{2}(\Omega_1-\Omega_2)\sin 2\theta+\lambda\cos 2\theta
\end{array}\right. .\end{equation}
It is legitimate, without loss of generality, to link $\theta$ to $\Omega_1$, $\Omega_2$ and $\lambda$ putting $\widetilde{K}$
into diagonal form: $\widetilde{\lambda}=0$, that is
\begin{equation}(\Omega_1-\Omega_2)\sin 2\theta=-2\lambda\cos 2\theta\label{10}\end{equation}
which may be satisfied whatever $\Omega_1$, $\Omega_2$ and $\lambda$ are.
To appreciate the role of the operators
$K$ and $S$ in our search of $V$, let's transform $\alpha_1$ and $\alpha_2$ using the unitary operator
$e^{iK}$\cite{172}
\begin{equation}\label{11}
    \beta_{\mu}\equiv e^{-iK}\alpha_{\mu}e^{iK}=S(\frac{\theta}{2})[S^{+}(\frac{\theta}{2})e^{-iK}S(\frac{\theta}{2})][S^{+}(\frac{\theta}{2})\alpha_{\mu}S(\frac{\theta}{2})][S^{+}(\frac{\theta}{2})e^{iK}S(\frac{\theta}{2})]S^{+}
\end{equation}
obtaining%

\begin{equation}\left\{\begin{array}{rl}
&\beta_1=(\cos^2\frac{\theta}{2}+e^{-i\eta}\sin^2\frac{\theta}{2})(e^{i\widetilde{\Omega}_1}\alpha_1)+\frac{1}{2}\sin\theta (e^{-i\eta}-1)(e^{i\widetilde{\Omega}_1}\alpha_2)
\\&  \beta_2=\frac{1}{2}\sin\theta (e^{-i\eta}-1)(e^{i\widetilde{\Omega}_1}\alpha_1)+(\sin^2\frac{\theta}{2}+e^{-i\eta}\cos^2\frac{\theta}{2})(e^{i\widetilde{\Omega}_1}\alpha_2)
\end{array}\right. \end{equation}
where $\eta=\widetilde{\Omega}_1-\widetilde{\Omega}_2$. 
Using finally the unitary operator $T=\exp[-i(\widetilde{\Omega}_1\alpha_1^{+}\alpha_1+(\widetilde{\Omega}_2+\pi)\alpha_2^{+}\alpha_2)]$
it is immediate to verify that 
\begin{equation}\left\{\begin{array}{rl}
&T^{+}\beta_1T=\widetilde{\alpha}_1
\\&T^{+}\beta_2T=\widetilde{\alpha}_2
\end{array}\right. \end{equation}
so that in conclusion we get that the unitary operator $V$ realizing the canonical transformation given by eq. (\ref{5}) may be taken in the form
\begin{equation}\label{14}
    V=e^{-i\pi\alpha_2^{+}\alpha_2}e^{-i\widetilde{\Omega}_1(\alpha_1^{+}\alpha_1+\alpha_2^{+}\alpha_2)}e^{iK}\equiv V(\theta,\eta)
\end{equation}
under the condition given by eq. (\ref{10}).
Now we concentrate on how $V$ transforms $H_{eff}$. To this end, we limit ourselves to notice that the
structure of $V^{+}H_{eff}V$ is the same as that of $H_{eff}$ for arbitrary values of the two parameters $\eta,\theta$ except that all the coefficients are now also $\eta-$ and $\theta-$ dependent. In view of our target we then search conditions for putting at zero the coefficients of the operators $\alpha_1^{+}\alpha_2$, $\alpha_2^{2}$, $\alpha_1\alpha_2$ (and of their h.c.) in the expression of $V^{+}H_{eff}V$. In this way we arrive to an algebraic system where we look for a compatible solution after imposing, as a nontrivial possibility, $\eta=\pi$. The resulting conditions become 
\begin{equation}\left\{\begin{array}{rl}
&\lambda_1\sin^2\theta+\lambda_2\cos^2\theta+\frac{1}{2}g\sin 2\theta=0
\\&(\lambda_1-\lambda_2)\sin 2\theta+g\cos2\theta=0
\\&\cos 2 \theta=0
\end{array}\right. .\end{equation}
then necessarily requiring  $\theta=\pi/4$, which in turn determines the model constraints  $\lambda_1=\lambda_2=\Lambda=-\frac{1}{2}g$. This solution is on the one hand compatible with eq. (\ref{10}), which indeed contributes only to establish the coefficients in $K$ and on the other hand reduces the canonical transformation given by eq. (\ref{5}) to the following one:

\begin{equation}\left\{\begin{array}{rl}
&\widetilde{\alpha}_1=\cos\frac{\pi}{4}\alpha_1+\sin\frac{\pi}{4}\alpha_2
\\&\widetilde{\alpha}_2=-\sin\frac{\pi}{4}\alpha_1+\cos\frac{\pi}{4}\alpha_2
\end{array}\right. \end{equation}
This transformation may be immediately accomplished by $S(\theta=\frac{\pi}{4})$ deducible from eq. (\ref{7}). The transformed hamiltonian then acquires the following form:

\begin{eqnarray} \label{17} V^{+}H_{eff}V=&\hbar\omega_0S_z+[\hbar\omega+(S-r_2)S_z-r_1]\alpha_1^{+}\alpha_1+[2\Lambda\alpha_1^{2}S_{+}+h.c.]\\\nonumber&+[\hbar\omega+(S+r_2)S_z+r_1]\alpha_2^{+}\alpha_2
\end{eqnarray}

The structure of $V^{+}H_{eff}V$ describes a fictious quadratic one mode-one atom generalized JCM and a free evolving collective mode, here labeled by
index $2$, whose presence influences in an effective way the energy separations of the true (that is untransformed) two-level system. Indeed, since $[\alpha_2^{+}\alpha_2,V^{+}H_{eff}V]=0$, we might study the dynamics fixing the Fock population of the decoupled mode. So doing, in the particular subspace wherein $\alpha_2^{+}\alpha_2$ assumes the integer nonnegative value $n_2$ with certainty, the dynamics of the decoupled quadratic Jaynes-Cummings mode is determined by the following one-atom--one-mode hamiltonian

\begin{eqnarray} \label{18} (V^{+}H_{eff}V)_{red}&=(\hbar\omega_0+n_2(s+r_2))S_z+[\hbar\omega+r_1]n_2+\\\nonumber&[2\Lambda\alpha_1^{2}S_{+}+h.c.]+[\hbar\omega+(S-r_2)S_z-r_1]\alpha_1^{+}\alpha_1
\end{eqnarray}
which exhibits an interesting \textquotedblleft double\textquotedblright intensity-dependent nature. In fact, the effective atomic energy separation turns out to be $n_2$-dependent as well as the effective energy coefficient of the coupled collective mode depends on $S_z$ \cite{post5}, \cite{post6}.

\section{Conclusive remarks}
Eqs. (\ref{17}), (\ref{18}) constitute the results of this short paper. The structure of $(V^{+}H_{eff}V)_{red}$ in particular provides in its own an interesting starting point to investigate on the physical origin of the parity effect and its consequences, without neglecting other possibly not yet explored quantum features characterizing the time evolution determined by the hamiltonian $H_{eff}$. For example, the property $[\alpha_2^{+}\alpha_2,V^{+}H_{eff}V]=0$ means that $H_{eff}$ has a constant of motion and the physical implications of this fact deserve to be investigated. A systematic study and possible further results will be presented elsewhere.
\ack
The authors gratefully acknowledge  by inspiring comments and stimulating discussions with A. Napoli and R. Messina. One of the authors(A.M.) acknowledges support by Italian Ministry of Research and Educations under project PRIN2008, responsible  prof. M.Cirillo, University Tor Vergata, Rome.

\end{document}